\documentclass[conference]{IEEEtran}

\usepackage[papersize={8.5in,11in},left=1.03in,right=1.03in,top=0.77in,bottom=1.03in]{geometry}
\usepackage{blindtext, graphicx}
\usepackage {tikz}
\usepackage{caption}
\usepackage{amssymb}
\usepackage{algpseudocode}
\usetikzlibrary {positioning}
\usepackage{tkz-euclide}
\usepackage{algcompatible,amsmath}
\usetkzobj{all}
\usepackage[T1]{fontenc}
\usepackage[utf8x]{inputenc}
\usetikzlibrary{calc}
\usepackage{amsthm}
\usepackage{thmtools}
\usepackage[linesnumbered,ruled]{algorithm2e}
\definecolor {processblue}{cmyk}{0.96,0,0,0}
\usepackage[absolute]{textpos}
\newcommand{\R}{\mathbb{R}}

\ifCLASSINFOpdf

\else

\fi

\begin{document}

\title{Scheduling Wireless Ad Hoc Networks in Polynomial Time Using Claw-free Conflict Graphs}

\author{
        \IEEEauthorblockN{Alper Köse\IEEEauthorrefmark{1}\IEEEauthorrefmark{2}, Muriel Médard\IEEEauthorrefmark{1}}
        \IEEEauthorblockA{
                \small
                \IEEEauthorrefmark{1}Research Laboratory of Electronics, Massachusetts Institute of Technology, \\ 77 Massachusetts Avenue, Cambridge, MA 02139, USA \\       
                \IEEEauthorrefmark{2}Department of Electrical Engineering, École Polytechnique Fédérale de Lausanne \\ Route Cantonale, 1015 Lausanne, Switzerland \\    
                \{akose, medard\}@mit.edu
                \vspace{0.0cm}
        }
    
}
\IEEEoverridecommandlockouts

\maketitle

\footnotetext[1]{This work has been presented at the 2017 IEEE PIMRC.}

\begin{abstract}

In this paper, we address the scheduling problem in wireless ad hoc networks by exploiting the computational advantage that comes when such scheduling problems can be represented by claw-free conflict graphs. It is possible to formulate a scheduling problem of network coded flows as finding maximum weighted independent set (MWIS) in the conflict graph of the network. We consider activation of hyperedges in a hypergraph to model a wireless broadcast medium. We show that the conflict graph of certain wireless ad hoc networks are claw-free. It is known that finding MWIS of a general graph is NP-hard, but in a claw-free conflict graph, it is possible to apply Minty's or Faenza et al.'s algorithms in polynomial time. We discuss our approach on some sample networks.

\end{abstract}

\begin{IEEEkeywords}
Wireless ad hoc networks; Claw-free graph; Conflict graph; Independent Set
\end{IEEEkeywords}

\IEEEpeerreviewmaketitle

\section{Introduction}

The problem of scheduling in wireless ad hoc networks is generally computationally expensive. In \cite{arikan1984some}, Arikan proves that scheduling is NP-complete in general packet radio networks when both primary and secondary interference are considered. In spread spectrum networks, complexity reduces as shown in \cite{hajek1988link}, where link-based scheduling can be done in polynomial time when each node can converse with at most one other node at a time. We schedule subsets of hyperedges in the hypergraph model of the network instead of doing classical link-based scheduling due to inability of link-based scheduling to capture the character of network coded flows. In \cite{ephremides1990scheduling}, authors prove the NP-completeness of scheduling broadcasts without tolerance of secondary conflicts. Another perspective for evaluation of interference is $K$-hop interference models in which no two links within $K$ hops can successfully transmit simultaneously. It is proved in \cite{sharma2006complexity} and \cite{sharma2006maximum} that link scheduling can be completed in polynomial time when $K=1$, otherwise it is NP-hard. In \cite{traskov2012scheduling}, scheduling of network coded flows with conflict graphs is considered, but without exploiting any graph characteristic. The model we illustrate is closely related to \cite{traskov2012scheduling}. Novelty of this paper comes from the fact that polynomial time scheduling property of wireless networks having claw-free conflict graphs is not investigated, and we contribute to literature by introducing some families of networks which can be scheduled in polynomial time. 

In a conflict graph of a network, each node symbolizes a valid transmission and an edge between two nodes represents that these two transmissions cannot be scheduled simultaneously, owing to interference constraints which can differ based on the assumptions on the network. The rate region of the network with random linear network coding is determined by three different constraints on capacity, flow and scheduling as in \cite{traskov2012scheduling}. Among these constraints, the determinant for optimization complexity is the scheduling constraint. Therefore, in this work, we concentrate on the scheduling constraint rather than the others. The scheduling problem can be formulated as finding maximum weighted independent set (MWIS) in the conflict graph where independent set is defined as follows. Given an undirected graph $\mathcal{G}=(\mathcal{V}, \mathcal{E})$, a subset of vertices $\mathcal{S}\subseteq \mathcal{V}$ is an independent set if $\{i,j\}\notin \mathcal{E}$ is satisfied for all $i$ and $j$ in $\mathcal{S}$. There are several examples for independent set formulation including the scheduling for routed traffic \cite{tassiulas1992stability} and network code construction in a wireline setup \cite{sundararajan2006systematic}. Finding MWIS dominates the complexity of the rate maximization problem in the rate region. MWIS problem is NP-hard for general graphs, therefore this makes scheduling NP-hard in general networks. On other hand, it is known that MWIS problem is solvable in polynomial time in claw-free graphs \cite{minty1980maximal}, \cite{nakamura2001revision}, \cite{faenza2014solving}. A graph $\mathcal{G}=(\mathcal{V}, \mathcal{E})$ is claw-free if none of its vertices $\mathcal{V}$ has three pairwise nonadjacent neighbours \cite{chudnovsky2005structure}.

The remainder of the paper is organized as follows. We detail on the conflict graph construction and present different scenarios, which are on line and tree networks for which the conflict graphs are claw-free, in Section II. As a next step, we explain the rate region constraints and discuss some results in a butterfly network in Section III. Finally, we conclude the paper in Section IV.

\section{Network Model and Conflict Graph}

Throughout the scenarios, we use the Protocol model \cite{gupta2000capacity} and K-hop interference model \cite{sharma2006complexity} with small variations to represent networks instead of the Physical model \cite{gupta2000capacity}, which takes $SINR$ levels into account.

\subsection{Scenario I}

We consider a wireless ad hoc network with $n$ nodes. We have a set of assumptions for the network:\\
A1.1) A node cannot receive from multiple nodes at the same time and a node cannot receive if it is in the interference range of another transmitting node. \\
A1.2) Time division duplex transceivers are used in network. \\
A1.3) The Protocol model \cite{gupta2000capacity} is used to decide interference relationships between transmissions. Suppose node $A$ is transmitting to node $B$ where node $C$ is simultaneously making a transmission to another node. Then, the transmission between $A$ and $B$ will be successful if and only if inequalities (1) and (2) are satisfied:

\begin{equation}
|A-B|\leq r_{T}
\end{equation}

\begin{equation}
|B-C|\geq (1+\Delta)|A-B|
\end{equation}

where $r_{T}$ is the maximum transmission range of nodes and $\Delta>0$ is the guard zone which can be chosen arbitrarily small for simplicity without loss of generality. Inequality (1) means that every node employs a common range of transmission $r_{T}$, and (2) says that in a communication pair, among all transmitting nodes, the receiver of this pair must be closest to its transmitter with a guard zone $\Delta$. 

Without making additional assumptions, the conflict graph of a network is not guaranteed to be claw-free.

\textbf{Construction 1 (Construction of Conflict Graph):} Reference \cite{traskov2012scheduling} shows the construction of conflict graph $\mathcal{G}=(\mathcal{V}, \mathcal{E})$ in a wireless network. Here, we use a similar strategy. It consists of the steps below:\\
C1.1) We can denote transmission $v$ as $(i,J)$ where $i$ is the source and $J$ as the set of receivers and $N(i)$ as the set of nodes that are in the transmission range of node $i$. The condition for a transmission being valid is given by $J\subset N(i)$, which means that the set of receiver nodes must be in the transmission range of the transmitter node. A valid transmission $v$ is modeled as a vertex, $v\in \mathcal{V}$. \\
C1.2) Say $v_{1},v_{2}\in \mathcal{V}$, then $\{v_{1},v_{2}\}\in \mathcal{E}$ if $v_{1}$ and $v_{2}$ cannot be scheduled simultaneously.\\
Let $v_{1}, v_{2}\in \mathcal{V}$ be in the conflict graph.$\{v_{1},v_{2}\}\in \mathcal{E}$ if any of the below conditions hold:\\
C1.2.1) $i_{1}=i_{2};$\\
C1.2.2) $(i_{1}\in J_{2}) || (i_{2}\in J_{1});$\\
C1.2.3) $J_{1}\cap J_{2}\neq \emptyset;$\\
C1.2.4) $|i_{2}-j|\leq (1+\Delta)|i_{1}-j|$ for $\exists j\in J_{1};$\\
C1.2.5) $|i_{1}-j|\leq (1+\Delta)|i_{2}-j|$ for $\exists j\in J_{2}$.

\textbf{Example 1:} Consider the topology given in Fig. 1. Assume a transmission range $r_{T}$ and an Euclidean distance $r$ where $r_{T}\geq r$ . In this case, when $A$ is a transmitting, for instance to $B$, then there will be interference to $(E,B), (F,C), (G,D)$ due to the Protocol model \cite{gupta2000capacity} and hence transmission node (A,B) shares edges with these three possible transmissions in the conflict graph. On the other hand, suppose $\angle BAC, \angle CAD, \angle DAB=120^{\circ}$, then there will not be any interference between any pairs of $(E,B), (F,C), (G,D)$ because these transmissions satisfy constraint (2) as well as (1). Consequently, this will lead to a claw in the conflict graph which can be seen in Fig. 2.

\begin{figure}
\begin{center}
\begin{tikzpicture}[scale=1,shorten >=1pt, auto, node distance=1cm,
   node_style/.style={scale=0.85,circle,draw=black,thick},
   edge_style/.style={draw=black, dashed}]

    \node[node_style] (v1) at (0,0) {$A$};
    \node[node_style] (v2) at (0,2) {$E$};
    \node[node_style] (v3) at (-1.732,-1) {$F$};
    \node[node_style] (v4) at (1.732,-1) {$G$};
    \node[node_style] (v5) at (0,1) {$B$};
    \node[node_style] (v6) at (-0.866,-0.5) {$C$};
    \node[node_style] (v7) at (0.866,-0.5) {$D$};
    
    \draw[edge_style]  (v1) edge node{$r$} (v5);
    \draw[edge_style]  (v5) edge node{$r$} (v2);
    \draw[edge_style]  (v1) edge node{$r$} (v6);
    \draw[edge_style]  (v6) edge node{$r$} (v3);
    \draw[edge_style]  (v1) edge node{$r$} (v7);
    \draw[edge_style]  (v7) edge node{$r$} (v4);
    
    \draw[->]  (v2) edge [bend left=30] (v5);
    \draw[->]  (v3) edge [bend left=30] (v6);
    \draw[->]  (v4) edge [bend left=30] (v7);
    \draw[->]  (v1) edge [bend right=30] (v5);
    
    \end{tikzpicture}
\end{center}
\caption{A wireless network example with the weights $r$ which show the Euclidean distance between nodes.}
\end{figure}
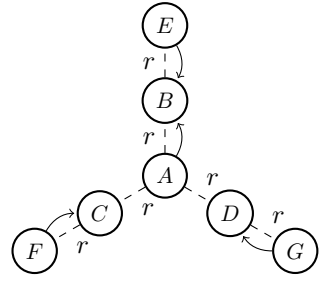

\begin{figure}
\begin{center}
\begin{tikzpicture}[scale=1,shorten >=1pt, auto, node distance=1cm,
   node_style/.style={scale=0.6,circle,draw=black,thick},
   edge_style/.style={draw=black}]

    \node[node_style] (v1) at (0,0) {$(A,B)$};
    \node[node_style] (v2) at (0,2) {$(E,B)$};
    \node[node_style] (v3) at (-1.732,-1) {$(F,C)$};
    \node[node_style] (v4) at (1.732,-1) {$(G,D)$};
    \draw[edge_style]  (v1) edge node{} (v2);
    \draw[edge_style]  (v1) edge node{} (v3);
    \draw[edge_style]  (v1) edge node{} (v4);
    \end{tikzpicture}
\end{center}
\caption{An induced subgraph of the conflict graph of the network in Fig. 1.}
\end{figure}
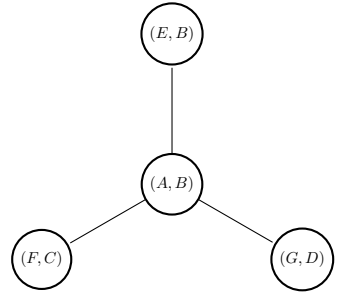

\textbf{Example 2:} A possible arrangement of wireless nodes to have a claw-free conflict graph is shown in Fig. 3. Source and sink can be thought as the nodes $A$ and $E$, respectively. Let the maximum possible transmission distance be $r_{T}$ and $\Delta$ be very small. Then, we can model the conflict graph of this network as seen in Fig. 4.  The independent set polytope is the convex hull of the incidence vectors of the five independent sets $\{(A,B),(D,E)\}$, $(A,C)$, $\{(B,C),(D,E)\}$, $(C,D)$ and $(A,\{B,C\})$.

\begin{figure}
\begin{center}
\begin{tikzpicture}[scale=0.85,shorten >=1pt, auto, node distance=1cm,
   node_style/.style={scale=0.85,circle,draw=black,thick},
   edge_style/.style={draw=black,dashed}]

    \node[node_style] (v1) at (0,0) {$A$};
    \node[node_style] (v2) at (2,0) {$B$};
    \node[node_style] (v3) at (3.5,0) {$C$};
    \node[node_style] (v4) at (6,0) {$D$};
    \node[node_style] (v5) at (8.5,0) {$E$};
    
    \draw[edge_style]  (v1) edge node{$2r_{T}/3$} (v2);
    \draw[edge_style]  (v2) edge node{$r_{T}/3$} (v3);
    \draw[edge_style]  (v3) edge node{$r_{T}$} (v4);
    \draw[edge_style]  (v4) edge node{$r_{T}$} (v5);
    
    \end{tikzpicture}
\end{center}
\caption{A possible physical arrangement of the wireless network which leads to a claw-free conflict graph for Scenario I.}
\end{figure}
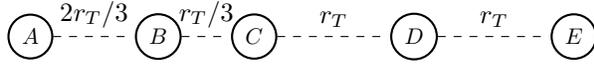

\begin{figure}
\begin{center}
\begin{tikzpicture}[scale=2.5,shorten >=1pt, auto, node distance=1cm,
   node_style/.style={scale=0.6,circle,draw=black,thick},
   edge_style/.style={draw=black}]

    \node[node_style] (v1) at (0,0) {$(A,B)$};
    \node[node_style] (v2) at (1,0) {$(A,C)$};
    \node[node_style] (v3) at (0,-1) {$(A,\{B,C\})$};
    \node[node_style] (v4) at (1,-1) {$(B,C)$};
    \node[node_style] (v5) at (-0.5,-0.5) {$(C,D)$};
    \node[node_style] (v6) at (1.5,-0.5) {$(D,E)$};

    \draw[edge_style]  (v1) edge node{} (v2);
    \draw[edge_style]  (v1) edge node{} (v3);
    \draw[edge_style]  (v1) edge node{} (v4);
    \draw[edge_style]  (v1) edge node{} (v5);
    \draw[edge_style]  (v2) edge node{} (v3);
    \draw[edge_style]  (v2) edge node{} (v4);
    \draw[edge_style]  (v2) edge node{} (v5);
    \draw[edge_style]  (v2) edge node{} (v6);
    \draw[edge_style]  (v3) edge node{} (v4);
    \draw[edge_style]  (v3) edge node{} (v5);
    \draw[edge_style]  (v3) edge node{} (v6);
    \draw[edge_style]  (v4) edge node{} (v5);
    \draw[edge_style]  (v5) edge node{} (v6);

    \end{tikzpicture}
\end{center}
\caption{Conflict graph of the network seen in Fig. 3.}
\end{figure}
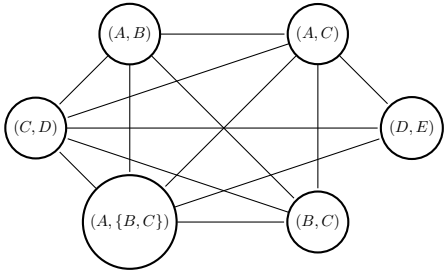

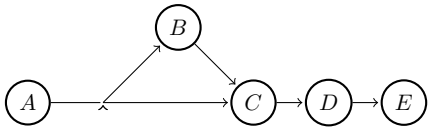
\begin{figure}
\begin{center}
\begin{tikzpicture}[scale=1,shorten >=1pt, auto, node distance=1cm,
   node_style/.style={scale=0.85,circle,draw=black,thick},
   edge_style/.style={draw=black}]

    \node[node_style] (v1) at (0,0) {$A$};
    \node[node_style] (v2) at (2,1) {$B$};
    \node[node_style] (v3) at (3,0) {$C$};
    \node[node_style] (v4) at (4,0) {$D$};
    \node[node_style] (v5) at (5,0) {$E$};

    \draw[edge_style]  (v1) edge node{} (1,0);
    \draw[->]  (1,0) edge node{} (v2);
    \draw[->]  (1,0) edge node{} (v3);
    \draw[->]  (v2) edge node{} (v3);
    \draw[->]  (v3) edge node{} (v4);
    \draw[->]  (v4) edge node{} (v5);
    
    \end{tikzpicture}
\end{center}
\caption{Hypergraph of the network seen in Fig. 3.}
\end{figure}

We can generalize Example 2 by realizing that the conflict graphs of the line networks are claw-free provided that there is enough distance between nodes to make 3 node away transmission impossible. We should physically satisfy $r_{T}<|\mu_{i}-\mu_{i+3}|$ for all $i\in \{1,2,...,n-3\}$ where $\mu_{i}$ is the $i$th node in the network. This inequality can be easily satisfied by many different positioning scenarios, so, for simplicity, we can use a hypergraph $\mathcal{H}=(\mathcal{N},\mathcal{A})$ to represent our model, where $\mathcal{N}$ denotes the nodes and $\mathcal{A}$ denotes the hyperedges to symbolize valid transmissions between wireless nodes. Hypergraph of a line network, which has claw-free conflict graph, changes depending on the number of nodes to which a node $i$ can transmit, which can be $1$ or $2$ for every $i\in \{1,2,...,n-2\}$ and $1$ for $i\in \{n-1\}$. We have $2^{n-2}$ different possible hypergraphs of a line network, with $n$ nodes, all of which lead to claw-free conflict graphs. For instance, hypergraph of the network in Fig. 3 can be seen in Fig. 5. Here, first node is able to transmit up to next two neighbours whereas the other nodes are only able to transmit to next node. Transmissions $(A,B)$ and $(C,D)$ are not simultaneously possible, because we use the Protocol model \cite{gupta2000capacity} to decide interference relations. Receiver $B$ is closer to $C$, a transmitter of another transmission, than to $A$ and this violates the constraint (2). Directed antennas, which we do not assume in our scenario, could be used to avoid the interference.

\textbf{Theorem 1:} \textit{Under the assumptions A1.1-3 and Construction 1, the conflict graph of a line network where a node is able to transmit to at most 2 nodes and all nodes convey information in the direction from $\mu_{1}$ to $\mu_{n}$ is guaranteed to be claw-free.}

\begin{figure}
\begin{center}
\begin{tikzpicture}[scale=1.4,shorten >=1pt, auto, node distance=1cm,
   node_style/.style={scale=0.85,circle,draw=black,thick},
   edge_style/.style={draw=black,dashed}]

    \node[node_style] (v1) at (0,0) {$A$};
    \node[node_style] (v2) at (1,0) {$B$};
    \node[node_style] (v3) at (2,0) {$C$};
    \node[node_style] (v4) at (3,0) {$D$};
    \node[node_style] (v5) at (4,0) {$E$};
    \node[node_style] (v6) at (5,0) {$F$};
    
    \draw[edge_style]  (v1) edge node{} (v2);
    \draw[edge_style]  (v2) edge node{} (v3);
    \draw[edge_style]  (v3) edge node{} (v4);
    \draw[edge_style]  (v4) edge node{} (v5);
    \draw[edge_style]  (v5) edge node{} (v6);
    \draw[->]  (v5) edge [bend right=30] (v6);
    \draw[->]  (v1) edge [bend right=30] (v2);
    \draw[->]  (v3) edge [bend right=30] (v4);
    
    \end{tikzpicture}
\end{center}
\caption{Illustration for proof of Theorem 1.}
\end{figure}
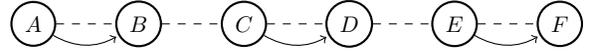

\begin{proof}
Let us prove this theorem by contradiction. To have a claw in the conflict graph, a transmission $v_{1}$ should have conflicts with three other transmissions $v_{2}, v_{3}, v_{4}$ whereas those three should not have any conflicts between them. Let us use Fig. 6 for ease of understanding. Since transmissions are from $\mu_{1}$ to $\mu_{n}$, from source to sink, assume that a node only transmits to nodes that are located closer to sink in terms of hop distance. To this end, assume $v_{1}$ as the central node of a possible claw, $v_{1}=(C,D)$. Then, we can have one interfering transmission from the source side of $C$, say $v_{2}=(A,B)$, and one from the sink side of $D$, $v_{3}=(E,F)$ not to have interference between $v_{2}$ and $v_{3}$. $B$ and $E$ are chosen to be as far as possible. In such situation, transmitting nodes $C$ and $E$ cause interference on receiving nodes $B$ and $D$, respectively. Now, we have to place the transmitter and the receiver of the last transmission. This one has to interfere with $(C,D)$ without interfering with $(A,B)$ and $(E,F)$ to induce a claw in the conflict graph of the network. If we place the transmitter on the source side of $C$, this leads to an interference with $(A,B)$ which will break the claw, so this option is not possible. Also, we cannot place the receiver in the sink side of $D$ since this leads to an interference with $(E,F)$ which will again break the claw. Therefore, since the receiver must be on the sink side relative to the transmitter, the only remaining option is to place both the transmitter and the receiver between $C$ and $D$. However, this option makes $C$ able to transmit to 3 different nodes where we assume each node is able to transmit to at most 2 nodes.
\end{proof}

\begin{figure}
\begin{center}
\begin{tikzpicture}[scale=1.05,shorten >=1pt, auto, node distance=1cm,
   node_style/.style={scale=0.8,circle,draw=black,thick},
   edge_style/.style={draw=black}]

    \node[node_style] (v1) at (0,0) {$v_{1}$};
    \node[node_style] (v2) at (1,0) {$v_{2}$};
    \node[node_style] (v3) at (5,0) {$v_{n-2}$};
    \node[node_style] (v4) at (6.5,0) {$v_{n-1}$};
    
    \node at ($(v2)!.5!(v3)$) {\ldots};
    
    \draw[edge_style]  (v1) edge node{} (v2);
    \draw[edge_style]  (v2) edge node{} (1.7,0);
    \draw[edge_style]  (4.1,0) edge node{} (v3);
    \draw[edge_style]  (v3) edge node{} (v4);
   
    \end{tikzpicture}
\end{center}
\caption{Conflict graph of a line network where only possible transmission for a node $i$ is $v_i=(i,i+1)$.}
\end{figure}
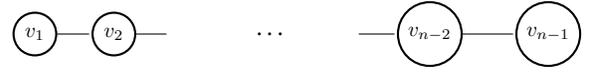

Let us consider a line network to see the effect of increasing number of nodes on end-to-end throughput, first when all transmissions are orthogonal and second when MWIS scheduling policy is used. To be very simple, suppose we have a line network with $n$ nodes and each node can only transmit to next node where node $\mu_{1}$ is the source and node $\mu_{n}$ is the sink. Suppose $n$ is odd. Also, assume that the transmission $(i,i+1)$ does not lead to an interference with $(i+2,i+3)$. We name the transmission $(i,i+1)$ as $v_{i}$. Under these assumptions, our conflict graph is shown in Fig. 7. So we have two different maximum independent sets, where weights are assumed equal, which are $S_{1}=\{v_{1},v_{3},...,v_{n-2}\}$ and $S_{2}=\{v_{2},v_{4},...,v_{n-1}\}$. Consider the situation where transmissions do not have any errors or loss. Since we only need 2 time slots to combine $S_{1}$ and $S_{2}$ to complete end-to-end transmissions, we conclude that end-to-end throughput is $1/2$ $bits/sec$ if we assume one time slot has a duration of 1 second and the link capacity between nodes is $1$ $bit/sec$. The throughput does not depend on $n$. However, if we assume all transmissions to be orthogonal, we can say that throughput will decrease linearly with the number of nodes in the network as it equals $1/(n-1)$ $bits/sec$ per frequency use. As seen, we have $0.5(n-1)$ gain on throughput by using MWIS approach. Using network coding does not provide any advantage in the end-to-end throughput of this system.

\subsection{Scenario II}
We can have other network topologies that induce a claw-free conflict graph. One of them is a tree representation of the network, but since it is harder to get claw-freeness with the same assumptions for the line topology model, we propose new set of assumptions:\\
A2.1) A node cannot receive from multiple nodes at the same time.\\
A2.2) Time division duplex transceivers are used in network. \\
A2.3) Nodes are arranged as a tree topology.\\
A2.4) We directly work on hypergraph model without any consideration on physical locations of nodes and assume an interference model based on hops instead of the Protocol model \cite{gupta2000capacity}.\\
A2.5) Transmissions are in the direction from root node to leaves of tree.\\
A2.6) Directed antennas are used, so interference can only occur in the forward direction along the tree.\\
A2.7) A transmitting node does not lead to any interference to the receivers which are 3-hops or more away from it.\\
A2.8) Only one node in every level can have children.

\textbf{Construction 2 (Construction of Conflict Graph):} In the construction of conflict graph, we implement the steps below:\\
C2.1) Model valid transmissions as the vertices $\mathcal{V}$ of conflict graph $\mathcal{G}=(\mathcal{V}, \mathcal{E})$ as in C1.1.\\
C2.2) Say $v_{1},v_{2}\in \mathcal{V}$, then $\{v_{1},v_{2}\}\in \mathcal{E}$ if $v_{1}$ and $v_{2}$ cannot be scheduled simultaneously.\\
Let $v_{1}, v_{2}\in \mathcal{V}$ be in the conflict graph.$\{v_{1},v_{2}\}\in \mathcal{E}$ if any of the below conditions hold:\\
C2.2.1) $i_{1}=i_{2};$\\
C2.2.2) $(i_{1}\in J_{2}) || (i_{2}\in J_{1});$\\
C2.2.3) $J_{1}\cap J_{2}\neq \emptyset;$\\
C2.2.4) ($i_{1}$ is a child of $i_{2}$) $||$ ($i_{2}$ is a child of $i_{1}$).

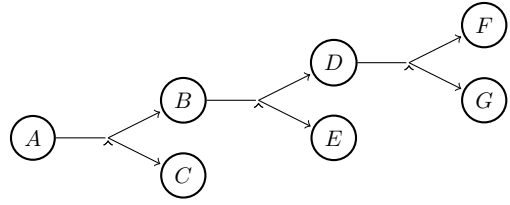
\begin{figure}
\begin{center}
\begin{tikzpicture}[scale=1,shorten >=1pt, auto, node distance=1cm,
   node_style/.style={scale=0.85,circle,draw=black,thick},
   edge_style/.style={draw=black}]

    \node[node_style] (v1) at (0,0) {$A$};   
    \node[node_style] (v2) at (2,0.5) {$B$};
    \node[node_style] (v3) at (2,-0.5) {$C$};    
    \node[node_style] (v4) at (4,1) {$D$};
    \node[node_style] (v5) at (4,0) {$E$};
    \node[node_style] (v6) at (6,1.5) {$F$};  
    \node[node_style] (v7) at (6,0.5) {$G$};
    
    \draw[edge_style]  (v1) edge node{} (1,0);
    \draw[->]  (1,0) edge node{} (v2);
    \draw[->]  (1,0) edge node{} (v3);
    \draw[edge_style]  (v2) edge node{} (3,0.5);
    \draw[->]  (3,0.5) edge node{} (v4);
    \draw[->]  (3,0.5) edge node{} (v5);
    \draw[edge_style]  (v4) edge node{} (5,1);
    \draw[->]  (5,1) edge node{} (v6);
    \draw[->]  (5,1) edge node{} (v7);

    \end{tikzpicture}
\end{center}
\caption{An example of a tree network which has claw-free conflict graph for Scenario II.}
\end{figure}

\begin{figure}
\begin{center}
\begin{tikzpicture}[scale=2.2,shorten >=1pt, auto, node distance=1cm,
   node_style/.style={scale=0.6,circle,draw=black,thick},
   edge_style/.style={draw=black}]

    \node[node_style] (v1) at (-1,1) {$(A,B)$};
    \node[node_style] (v2) at (-1,0) {$(A,C)$};
    \node[node_style] (v3) at (-1,-1) {$(A,\{B,C\})$};
    \node[node_style] (v4) at (0,1) {$(B,D)$};
    \node[node_style] (v5) at (0,0) {$(B,E)$};
    \node[node_style] (v6) at (0,-1) {$(B,\{D,E\})$};
    \node[node_style] (v7) at (1,1) {$(D,F)$};
    \node[node_style] (v8) at (1,0) {$(D,G)$};
    \node[node_style] (v9) at (1,-1) {$(D,\{F,G\})$};
 
    \draw[edge_style]  (v1) edge node{} (v2);
    \draw[edge_style]  (v1) edge [bend right=20] (v3);
    \draw[edge_style]  (v2) edge node{} (v3);
    \draw[edge_style]  (v4) edge node{} (v5);
    \draw[edge_style]  (v4) edge [bend right=20] (v6);
    \draw[edge_style]  (v5) edge node{} (v6);
    \draw[edge_style]  (v7) edge node{} (v8);
    \draw[edge_style]  (v7) edge [bend left=20] (v9);
    \draw[edge_style]  (v8) edge node{} (v9);
    \draw[edge_style]  (v1) edge node{} (v4);
    \draw[edge_style]  (v1) edge node{} (v5);
    \draw[edge_style]  (v1) edge node{} (v6);
    \draw[edge_style]  (v2) edge node{} (v4);
    \draw[edge_style]  (v2) edge node{} (v5);
    \draw[edge_style]  (v2) edge node{} (v6);
    \draw[edge_style]  (v3) edge node{} (v4);
    \draw[edge_style]  (v3) edge node{} (v5);
    \draw[edge_style]  (v3) edge node{} (v6);
    \draw[edge_style]  (v4) edge node{} (v7);
    \draw[edge_style]  (v4) edge node{} (v8);
    \draw[edge_style]  (v4) edge node{} (v9);
    \draw[edge_style]  (v5) edge node{} (v7);
    \draw[edge_style]  (v5) edge node{} (v8);
    \draw[edge_style]  (v5) edge node{} (v9);
    \draw[edge_style]  (v6) edge node{} (v7);
    \draw[edge_style]  (v6) edge node{} (v8);
    \draw[edge_style]  (v6) edge node{} (v9);

    \end{tikzpicture}
\end{center}
\caption{Conflict graph of the network seen in Fig. 8.}
\end{figure}
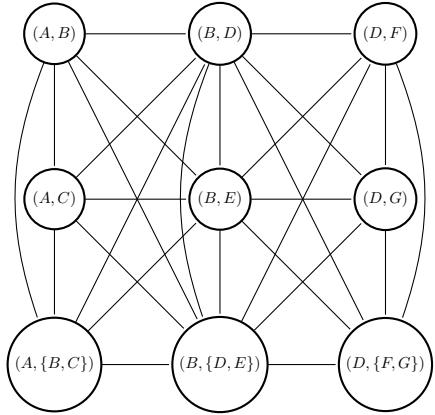

\textbf{Theorem 2:} \textit{Under the assumptions A2.1-8 and Construction 2, conflict graph of a wireless network is guaranteed to be claw-free.}

\begin{proof}
Let us assume that the root node belongs to level $1$ and a node, which has a distance $k$ to root in terms of hyperedge number, belongs to level $k+1$. Now, consider the transmission $v_{1}$, from a node $i_{k}$ in level $k$ to its children that reside in level $k+1$. We have $2^{N_{k}}-1$ different interfering transmissions to $v_{1}$ which are from level $k-1$ to $k$ where $N_{k}$ is the number of children of $i_{k}$'s parent. Since only one of these transmissions can be scheduled in one time slot, they induce a complete subgraph in the conflict graph. Therefore, we can only choose one transmission, say $v_{2}$, from level $k-1$ to $k$ which has interference with $v_{1}$ because the cardinality of maximum independent set of the complete graph is $1$. Assuming that we have the node $i_{k+1}$, which belongs to level $k+1$ and has children, in the receiver set of the transmission $v_{1}$ (otherwise, it is easier to say that we will not have a claw.), we have another complete subgraph which contains the transmissions from the node $i_{k+1}$ to its children in level $k+2$. One of these transmissions can be selected for a possible claw, say $v_{3}$. Since, it is not possible to find another independent transmission $v_{4}$, we have a claw-free conflict graph for the network. 
\end{proof}

An example of a tree network which has a claw-free conflict graph and its conflict graph can be seen in Fig. 8 and Fig. 9, respectively. 

\subsection{Scenario III}
In this scenario, transmission and interference schemes are less restricted, since we only allow for full-duplex transceivers and do not allow for 2-hop interference, compared to Scenario II, but the network topology is relaxed by letting every node have children. Assumptions are below:\\
A3.1) A node cannot receive from multiple nodes at the same time.\\
A3.2) Full duplex transceivers are used in network.\\
A3.3) Nodes are arranged as a tree topology.\\
A3.4) We directly work on hypergraph model without any consideration on physical locations of nodes and assume an interference model based on hops instead of the Protocol model \cite{gupta2000capacity}.\\
A3.5) Transmissions are in the direction from root node to leaves of tree.\\
A3.6) Directed antennas are used, so interference can only occur in the forward direction along the tree.\\
A3.7) A transmitting node does not lead to any interference to the receivers which are 2-hops or more away from it.

\textbf{Construction 3 (Construction of Conflict Graph):} In the construction of conflict graph, we implement the steps below:\\
C3.1) Model valid transmissions as the vertices $\mathcal{V}$ of conflict graph $\mathcal{G}=(\mathcal{V}, \mathcal{E})$ as in C1.1.\\
C3.2) Say $v_{1},v_{2}\in \mathcal{V}$, then $\{v_{1},v_{2}\}\in \mathcal{E}$ if $v_{1}$ and $v_{2}$ cannot be scheduled simultaneously.\\
Let $v_{1}, v_{2}\in \mathcal{V}$ be in the conflict graph.$\{v_{1},v_{2}\}\in \mathcal{E}$ if any of the below conditions hold:\\
C3.2.1) $i_{1}=i_{2};$\\
C3.2.2) $J_{1}\cap J_{2}\neq \emptyset;$\\

\textbf{Theorem 3:} \textit{Under the assumptions A3.1-7 and Construction 3, conflict graph of a wireless network is guaranteed to be claw-free.}

\begin{proof}
Let us again assume that the root node belongs to level $1$ and a node, which has a distance $k$ to root in terms of hyperedge number, belongs to level $k+1$. In these settings, we eliminate the possibility of having interference between any transmissions from different levels. Let us denote a transmission from a node in level $k$ to a node in level $k+1$ as $k\rightarrow k+1$. Any transmission $k\rightarrow k+1$ interferes neither with $k-1\rightarrow k$ nor with $k+1\rightarrow k+2$ because we use full-duplex transceivers which makes a node eligible to transmit and receive at the same time. Therefore, we only have interference between the transmissions which initiate from the same node. A node with $j$ children has $2^{j}-1$ possible transmissions where each pair has interference between them which then leads to a complete graph with $2^{j}-1$ nodes in the conflict graph. Assume we have $m$ non-leaf tree nodes, then we have $m$ disjoint complete subgraphs in the overall conflict graph which implies that the conflict graph is claw-free.
\end{proof}
 
An example of a tree network which has a claw-free conflict graph and its conflict graph can be seen in Fig. 10 and Fig. 11, respectively. 

\begin{figure}
\begin{center}
\begin{tikzpicture}[scale=1,shorten >=1pt, auto, node distance=1cm,
   node_style/.style={scale=0.85,circle,draw=black,thick},
   edge_style/.style={draw=black}]

    \node[node_style] (v1) at (0,0) {$A$};   
    \node[node_style] (v2) at (2,0.5) {$B$};
    \node[node_style] (v3) at (2,-0.5) {$C$};    
    \node[node_style] (v4) at (4,1.5) {$D$};
    \node[node_style] (v5) at (4,-0.5) {$E$};
    \node[node_style] (v6) at (6,2) {$F$};  
    \node[node_style] (v7) at (6,1) {$G$};
    \node[node_style] (v8) at (6,0) {$H$};  
    \node[node_style] (v9) at (6,-1) {$I$};
    
    \draw[edge_style]  (v1) edge node{} (1,0);
    \draw[->]  (1,0) edge node{} (v2);
    \draw[->]  (1,0) edge node{} (v3);
    \draw[edge_style]  (v2) edge node{} (3,0.5);
    \draw[->]  (3,0.5) edge node{} (v4);
    \draw[->]  (3,0.5) edge node{} (v5);
    \draw[edge_style]  (v4) edge node{} (5,1.5);
    \draw[->]  (5,1.5) edge node{} (v6);
    \draw[->]  (5,1.5) edge node{} (v7);
    \draw[edge_style]  (v5) edge node{} (5,-0.5);
    \draw[->]  (5,-0.5) edge node{} (v8);
    \draw[->]  (5,-0.5) edge node{} (v9);
    
    \end{tikzpicture}
\end{center}
\caption{An example of a tree network which has a claw-free conflict graph for Scenario III but not for Scenario II.}
\end{figure}
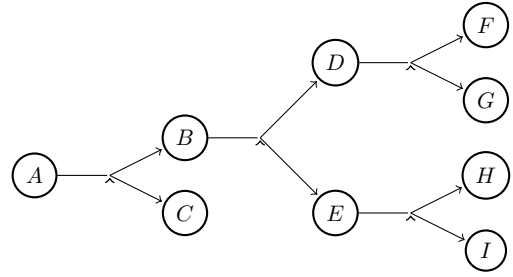

\begin{figure}
\begin{center}
\begin{tikzpicture}[scale=1.5,shorten >=1pt, auto, node distance=1cm,
   node_style/.style={scale=0.6,circle,draw=black,thick},
   edge_style/.style={draw=black}]

    \node[node_style] (v1) at (-1,1) {$(A,B)$};
    \node[node_style] (v2) at (-1,0) {$(A,C)$};
    \node[node_style] (v3) at (-1,-1) {$(A,\{B,C\})$};
    \node[node_style] (v4) at (0,1) {$(B,D)$};
    \node[node_style] (v5) at (0,0) {$(B,E)$};
    \node[node_style] (v6) at (0,-1) {$(B,\{D,E\})$};
    \node[node_style] (v7) at (1,1) {$(D,F)$};
    \node[node_style] (v8) at (1,0) {$(D,G)$};
    \node[node_style] (v9) at (1,-1) {$(D,\{F,G\})$};
    \node[node_style] (v10) at (2,1) {$(E,H)$};
    \node[node_style] (v11) at (2,0) {$(E,I)$};
    \node[node_style] (v12) at (2,-1) {$(E,\{H,I\})$};

    \draw[edge_style]  (v1) edge node{} (v2);
    \draw[edge_style]  (v1) edge [bend right=30] (v3);
    \draw[edge_style]  (v2) edge node{} (v3);
    \draw[edge_style]  (v4) edge node{} (v5);
    \draw[edge_style]  (v4) edge [bend right=30] (v6);
    \draw[edge_style]  (v5) edge node{} (v6);
    \draw[edge_style]  (v7) edge node{} (v8);
    \draw[edge_style]  (v7) edge [bend left=30] (v9);
    \draw[edge_style]  (v8) edge node{} (v9);
    \draw[edge_style]  (v10) edge node{} (v11);
    \draw[edge_style]  (v10) edge [bend left=30] (v12);
    \draw[edge_style]  (v11) edge node{} (v12);

    \end{tikzpicture}
\end{center}
\caption{Conflict graph of the network seen in Fig. 10.}
\end{figure}
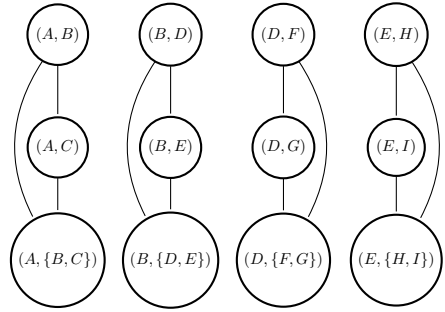

\section{Rate Region}

\subsection{Rate Region Constraints}

To find the rate region, we assume that we have a hypergraph $\mathcal{H}=(\mathcal{N},\mathcal{A})$ model of the network as in Fig. 8 where $\mathcal{N}$ symbolizes the nodes and $\mathcal{A}$ symbolizes the hyperedges. We consider the multicast connection with rate $R$, we have a source $s\in \mathcal{N}$ and sinks $\mathcal{T}\in \mathcal{N}$ where all sinks $\mathcal{T}\in \mathcal{N}$ demand the same information and network coding is used. We can extend it to multiple connections using intra-session coding. When we use $i$ to denote a wireless node, $N(i)$ denotes the set of nodes to which node $i$ can transmit, namely there is a hyperedge from $i$ to set $N(i)$. Recall that MWIS problem dominates the complexity of the rate maximization problem in the rate region. In this section, we review the three constraints of the rate region to highlight the importance of the scheduling problem that we address by revisiting \cite{traskov2012scheduling} and \cite{lun2006minimum} and give the formulation of MWIS problem for the conflict graph.

Firstly, we have \textit{capacity constraints} given in (3).

\begin{equation}
\sum_{j\in K}^{} x^{(t)}_{ij}\leq \sum_{j\subset N(i)}^{} z_{ij}
\end{equation}

$\forall i \in \mathcal{N}, K\subset N(i), t \in \mathcal{T}.$

In constraint (3), $x^{(t)}_{ij}$ shows the flow rate of data packets from node $i$ to node $j$ in the way of sink $t$. $z_{iJ}$ is the average packet injection rate to the output set $J$ of hyperedge $(i,J)$ and $z_{iJ}$ is guaranteed to exist and be finite if the injection process is assumed to be stationary and ergodic.

Second, we have \textit{flow constraints} given in (4) and (5). 

\begin{equation}
\sum_{j\in N(i)}^{} x^{(t)}_{ij} - \sum_{\{j|i\in N(j)\}}^{} x^{(t)}_{ji} = 
\begin{cases}
      R, & \text{if}\ i=s \\
      -R, & \text{if}\ i=t \\
      0, & \text{otherwise}
    \end{cases}
\end{equation}

$\forall i \in \mathcal{N}, t \in \mathcal{T}.$

\begin{equation}
x^{(t)}_{ij}\geq 0
\end{equation}

$\forall i \in \mathcal{N}, j\in N(i), t \in \mathcal{T}.$

Constraint (4) should be satisfied to guarantee a flow of rate $R$ from the source to sinks and it is trivial that flow rates cannot be negative as seen in (5).

Lastly, we have the \textit{scheduling constraint} given in (6).

\begin{equation}
\mathbf{z}\in P_{IND}(\mathcal{G}).
\end{equation}

In (6), vector $\mathbf{z}=(z_{iJ})$ is called network coding subgraph where $(i,J)\in \mathcal{A}$. $P_{IND}(\mathcal{G})$ denotes the independent set polytope of $\mathcal{G}=(\mathcal{V}, \mathcal{E})$, which is the conflict graph of the considered network. In overall, the network coding subgraph should lie in the independent set polytope of the conflict graph.

The maximum weighted independent set problem of a weighted undirected graph $\mathcal{G}=(\mathcal{V}, w, \mathcal{E})$ with $n$ vertices where $w:\mathcal{V}(\mathcal{G})\rightarrow \R$ can be formulated as below:

Maximize $\sum_{i\in \mathcal{V}}^{} w_{i}v_{i} $

Subject to $v_{i}+v_{j}\leq 1 \quad \{i,j\}\in \mathcal{E}$

where $v_{i}\in \{0,1\}$  for $1\leq i\leq n$.

We maximize $R$ subject to (3)-(6). Scheduling constraint (6) can be formulated as MWIS problem, and it is known to be NP-hard for general graphs to solve which makes our optimization problem NP-hard. However, there are algorithms proposed for solving MWIS problem in claw-free graphs in polynomial time. For instance, Minty's \cite{minty1980maximal}, \cite{nakamura2001revision} algorithm can be used for solvability in $O(n^{6})$ or Faenza et al.'s \cite{faenza2014solving} algorithm can be used for further improvement to $O(n^{3})$. Since we are able to find MWIS of the claw-free conflict graph in polynomial time, it is possible to solve the joint subgraph optimization and scheduling problem, i.e. maximizing R subject to (3)-(6), in polynomial time \cite{traskov2008scheduling}.

\subsection{Illustration}

Networks having claw-free conflict graphs are not limited with the discussed ones in Section II. Let us consider a butterfly network as seen in Fig. 12 with the assumptions A1.1-3 which lead to Construction 1 where $r\leq r_{T}$, $|AD|> r_{T}$ and $\Delta$ is negligible. $A$ is the source node where $E$ and $F$ are the sinks of a multicast connection. First, we show that butterfly network has claw-free conflict graph. There are $12$ valid transmissions, $3$ for each node being source except $E$ and $F$. Therefore, possible transmissions from a node lead to a complete subgraph of 3 nodes. To have a claw, a transmission $(i_{1},J_{1})$ should have three conflicts, each with a different source $i_{a}\neq i_{b}$ for $a,b \in \{1,2,3,4\}$ where these three should not have any conflicts between them. However, this is not possible in the given butterfly network which lead to a claw-free conflict graph. So, the scheduling can be completed in polynomial time. Complement of the conflict graph can be seen in Fig. 13 for a more clear illustration.

Assume all transmissions are done without any errors or loss from source to sinks and every link has a unit capacity, say $1$ $bit/sec$. We have $2$ bits to send in the multicast connection. We consider four different conditions where the constraints (3)-(6) from the previous subsection only apply to MWIS approach with network coding:\\
1) All transmissions are orthogonal without network coding.\\
2) All transmissions are orthogonal with network coding.\\
3) MWIS approach without network coding.\\
4) MWIS approach with network coding.

In the fully orthogonal model \cite{lun2006minimum}, all transmission signals are orthogonal, making the network interference-free but this leads to a suboptimal bandwith usage. We maximize bandwith efficiency by frequency reuse whenever possible. We could also consider the two-hop constraints model \cite{traskov2012scheduling} in which if a node $i$ transmits, none of the nodes in two-hop neighbourhood of $i$ can transmit, but it adds up to the same thing with orthogonal model in this butterfly network. In 1-3, our conflict graph construction, Construction 1, does not represent the conditions correctly owing to some additional constraints that show up. For instance, when we adopt orthogonal model, conflict graph becomes a complete graph where we can choose only one transmission for a time slot due to our bandwith limited regime. Our conflict graph formulation can only be used for MWIS approach with network coding. Throughput analysis can be seen below:\\
1) We can achieve an end-to-end throughput of $1/3$ $bits/sec$ per frequency use.\\
2) We can achieve an end-to-end throughput of $2/5$ $bits/sec$ per frequency use which is better than the first approach.\\
3) We are now scheduling $\{(A,B),(C,\{D,F\})\}$, $\{(A,C),(B,\{D,E\})\}$, $(D,E)$ and $(D,F)$ in four time slots, thus have an end-to-end throughput of $1/2$ $bits/sec$ which is better than second approach.\\
4) We first set the conflict graph as seen in Fig. 13, then find the maximum weighted independent sets, and finally use convex combination of them for scheduling. We have $\{(A,B),(C,\{D,F\})\}$, $\{(A,C),(B,\{D,E\})\}$ and $(D,\{E,F\})$ which we can combine in three time slots. End-to-end throughput is $2/3$ $bits/sec$ which is the highest among four conditions.

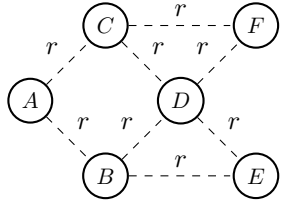
\begin{figure}
\begin{center}
\begin{tikzpicture}[scale=1,shorten >=1pt, auto, node distance=1cm,
   node_style/.style={scale=0.85,circle,draw=black,thick},
   edge_style/.style={draw=black, dashed}]

    \node[node_style] (v1) at (0,0) {$A$};
    \node[node_style] (v2) at (1,-1) {$B$};
    \node[node_style] (v3) at (1,1) {$C$};
    \node[node_style] (v4) at (2,0) {$D$};
    \node[node_style] (v5) at (3,-1) {$E$};
    \node[node_style] (v6) at (3,1) {$F$};

    \draw[edge_style]  (v1) edge node{$r$} (v2);
    \draw[edge_style]  (v1) edge node{$r$} (v3);
    \draw[edge_style]  (v2) edge node{$r$} (v4);
    \draw[edge_style]  (v2) edge node{$r$} (v5);
    \draw[edge_style]  (v3) edge node{$r$} (v4);
    \draw[edge_style]  (v3) edge node{$r$} (v6);
    \draw[edge_style]  (v4) edge node{$r$} (v5);
    \draw[edge_style]  (v4) edge node{$r$} (v6);

    \end{tikzpicture}
\end{center}
\caption{A butterfly network with the weights $r$ which show the Euclidean distance between nodes.}
\end{figure}

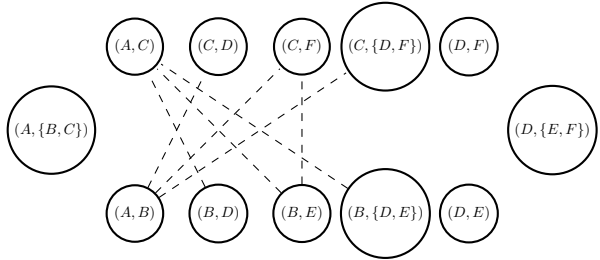
\begin{figure}
\begin{center}
\begin{tikzpicture}[scale=1.11,shorten >=1pt, auto, node distance=1cm,
   node_style/.style={scale=0.56,circle,draw=black,thick},
   edge_style/.style={draw=black,dashed}]

    \node[node_style] (v1) at (0,0) {$(A,\{B,C\})$};
    \node[node_style] (v2) at (1,-1) {$(A,B)$};
    \node[node_style] (v3) at (2,-1) {$(B,D)$};
    \node[node_style] (v4) at (3,-1) {$(B,E)$};
    \node[node_style] (v5) at (4,-1) {$(B,\{D,E\})$};
    \node[node_style] (v6) at (5,-1) {$(D,E)$};   
    \node[node_style] (v7) at (6,0) {$(D,\{E,F\})$};
    \node[node_style] (v8) at (5,1) {$(D,F)$};
    \node[node_style] (v9) at (4,1) {$(C,\{D,F\})$};
    \node[node_style] (v10) at (3,1) {$(C,F)$};
    \node[node_style] (v11) at (2,1) {$(C,D)$};
    \node[node_style] (v12) at (1,1) {$(A,C)$};

    \draw[edge_style]  (v2) edge node{} (v9);
    \draw[edge_style]  (v2) edge node{} (v10);
    \draw[edge_style]  (v2) edge node{} (v11);    
    \draw[edge_style]  (v3) edge node{} (v12);
    \draw[edge_style]  (v4) edge node{} (v10);
    \draw[edge_style]  (v4) edge node{} (v12);    
    \draw[edge_style]  (v5) edge node{} (v12);

    \end{tikzpicture}
\end{center}
\caption{Complement of the conflict graph of the network seen in Fig. 12.}
\end{figure}

\section{Conclusion}

We have investigated wireless ad hoc networks which have claw-free conflict graphs. To this end, we used conflict graph to model possible valid transmissions and their interference relations in the network, then we exploited the claw-freeness of the conflict graph of certain networks and MWIS algorithm for graphs to do the scheduling in polynomial time. We believe that this paper can be used as a guide to set up wireless networks which allow jointly solving the network coding subgraph and the scheduling problem in polynomial time. However, most of the real life networks do not have claw-free conflict graphs and our approach can only be applied to limited number of networks.

For future work, we work on a strategy which breaks all of the claws in a conflict graph by iteratively introducing edges. We believe this new strategy will allow us to do polynomial time scheduling for more general networks with a very small throughput loss. Besides this new strategy, the Physical model \cite{gupta2000capacity} can be considered to form the conflict graph of networks which may complicate scenarios but give a more detailed view at the same time. Lastly, a superset of claw-free graphs in which MWIS problem can be solved in polynomial time can be searched to extend throughput optimal polynomial time scheduling to more general networks.

\ifCLASSOPTIONcaptionsoff
  \newpage
\fi

\bibliographystyle{IEEEtran}  
\bibliography{references}

\end{document}